\documentclass[prl,twocolumn,showpacs,preprintnumbers,amsmath,amssymb]{revtex4}

\usepackage{graphicx}
\usepackage{dcolumn}
\usepackage{bm}

\begin{document}

\title{Kelvin waves of quantized vortex lines in trapped Bose-Einstein condensates}
\author{T. P. Simula, T. Mizushima, and K. Machida}
\affiliation{Department of Physics, Okayama University, Okayama 700-8530, Japan}

\begin{abstract}
We have theoretically investigated Kelvin waves of quantized vortex lines in trapped Bose-Einstein condensates. Counter-rotating perturbation induces an elliptical instability to the initially straight vortex line, driven by a parametric resonance between a quadrupole mode and a pair of Kelvin modes of opposite momenta. Subsequently Kelvin waves rapidly decay to longer wavelengths emitting sound waves in the process. We present a modified Kelvin wave dispersion relation for trapped superfluids and propose a simple method to excite Kelvin waves of specific wave number.
\end{abstract}

\maketitle
In classical hydrodynamics a vortex line is the trajectory obtained by following the local vorticity vector, defined as the curl of the velocity field of the fluid \cite{Donnelly1991a}. Wing tip vortices of aircrafts, tornadoes, whirlpools in a flowing water, and cosmic strings provide intuitive mental pictures of vortices. They support a branch of chiral normal modes in which the perturbation propagates along the vortex line and rotates about its unperturbed position therefore distorting the vortex into a helical shape. In 1880, Thomson (Lord Kelvin) derived the dispersion relation for such excitation modes, which are now known as Kelvin waves \cite{Thomson1880a}. These helically vibrating normal modes are objects of fundamental importance in classical hydrodynamics. 

Similar structures also exist in superfluids in which vorticity is quantized in units of $h/m$, where $h$ is  Planck's constant and $m$ is the mass of the superfluid particle \cite{Pitaevskii1961a, Sonin1987a, Donnelly1991a, Epstein1992a, Fetter2004a, Martikainen2004a}. Kelvin modes or kelvons of singly quantized rectilinear superfluid vortices have no radial nodes and in axisymmetric systems they are characterized by their angular momentum quantum number $\ell =-1$ and momentum $\hbar{\bf k}_z$ along the vortex axis. First experimental evidence of Kelvin waves in superfluids was obtained by Hall and they are thought to play an essential role in the formation and decay of superfluid turbulence \cite{Hall1958a}. In particular, Kelvin wave and Kolmogorov-type cascades are predicted to mediate the energy transfer across length scales in turbulent superfluid systems \cite{Svistunov1995a, Vinen2001a, Leadbeater2003a,Walmsley2007a, Kozik2008a}. Bretin \emph{et al.} were able to excite and observe Kelvin waves in a single quantum vortex in Bose-Einstein condensates \cite{Bretin2003a}. Their experimental kelvon excitation mechanism was theoretically clarified by Mizushima \emph{et al.} \cite{Mizushima2003a, Mizushima2004b}. The ability to address individual vortex lines in these systems opens the possibility to study Kelvin waves and their relation to elementary processes governing superfluid turbulence.

In this Letter we study the microscopic dynamics of Kelvin waves of quantized vortex lines in Bose-Einstein condensates by employing direct simulations of the three-dimensional time-dependent Gross-Pitaevskii equation combined with an analysis of the collective Bogoliubov excitations of the unperturbed state. We treat the dynamics of Kelvin waves fully quantum mechanically in contrast to previously employed phenomenological models such as in Refs.\cite{Schwarz1985a,Vinen2003a}. We find counter-rotating quadrupole surface mode (surfon) excitations to decay via Beliaev mechanism into a pair of kelvons of opposite momenta in perfect agreement with Refs.\cite{Bretin2003a, Mizushima2003a, Mizushima2004b}. Subsequently the primary kelvons are found to decay into kelvons of longer wavelength by emitting sound waves in the scattering process. This mechanism may help to explain microscopically how superfluid turbulence returns to laminar flow even at zero temperature. We introduce a modification to the semi-classical kelvon dispersion relation extending its validity to trapped superfluids. Finally, we propose an experimentally feasible method to excite kelvons of particular wave number.  

\begin{figure}
\includegraphics[width=1\columnwidth]{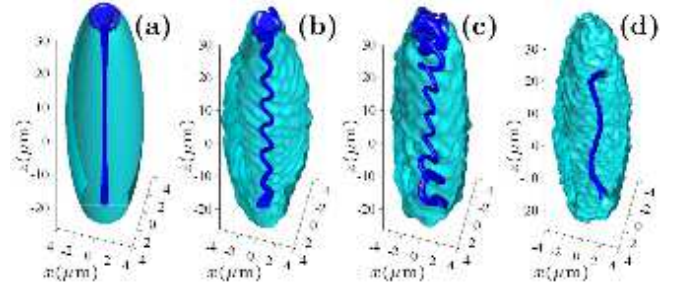}
\caption{(Color online) Kelvin waves in a quantum mechanical vortex. The initially straight vortex (a) undergoes an elliptical instability and the vortex line shape changes from plane-sinusoidal (b) to helical (c). Subsequently the primary kelvons decay into kelvons of longer wavelength (d) emitting sound waves seen as ripples at the condensate surface. In the laboratory frame the sense of the vortex motion changes sign between frames (c) and (d). The frames are for times 0 ms (a) 103 ms (b) 129 ms (c), and 350 ms (d). The bottom end-cap of each frame has been cut-off for the aid of visualization.}
\label{fig1}
\end{figure}

Following the experimental procedure of Bretin \emph{et al.} \cite{Bretin2003a} we consider a Bose-Einstein condensate composed of $1.3\times10^5$ $^{87}$Rb particles trapped in a harmonic potential whose transverse and axial frequencies are $\omega_\perp=2\pi\times 98.5$ Hz and $\omega_z=2\pi\times11.8$ Hz, respectively. We model the physics of such system by numerically solving the pure (no additional damping terms included) time-dependent Gross-Pitaevskii equation on a parallel computer using the method described in Refs.\cite{Schneider, Simula2008a}.  We complement the dynamical simulations by directly solving the Bogoliubov-de Gennes eigenvalue equations obtaining the full Kelvin wave dispersion relation \cite{Mizushima2003a, Mizushima2004b}. Our initial condition is a near-axisymmetric single-quantum vortex which possess one unit of angular momentum per particle. Such state, shown in Fig.\ref{fig1}(a), is a local energy minimum in the non-rotating trap rather than an absolute ground state which would have zero vorticity. 

At $t=0$ we suddenly switch on an elliptical rotating perturbation
\begin{equation}
V_{\rm pert}({\bf r},t)=\frac{1}{2}\epsilon m \omega^2_\perp [   ( x^2 - y^2) \cos (2\Omega t)+ 2 xy\sin(2\Omega t)  ],
\label{eq1}
\end{equation}
where $\epsilon=0.025$, and $m$ is the mass of the atom. The frequency $\Omega=-0.6 \omega_\perp$ is resonant with the counter-rotating (with respect to the superfluid flow) $\ell=-2$ quadrupole surfon of the unperturbed condensate \cite{Bretin2003a}. This resonant external drive transfers population from the initial vortex state to the counter-rotating quadrupole surfon state thus changing the total orbital angular momentum of the system. When the perturbation is switched off at $t=35$ms, the $z$-component of the angular momentum has been reduced from $L_z/N=1\hbar$ to $0.47\hbar$. From there on $L_z$ is a constant of motion since the bare trap potential is cylindrically symmetric. 
In contrast, if the system is continuously rotated, the growing surface instability leads to (anti)vortex nucleation as in nonrotating systems \cite{Isoshima1999a, Madison2001a, Sinha2001a, Simula2002a, Penckwitt2002a} and eventually Lz changes its sign. If $\Omega$ is held constant the angular momentum transfer is limited since the rotating perturbation falls out of resonance when the quadrupole surfon becomes frequency shifted due to the changing angular momentum.

The vortex line stays straight until $t=64$ ms at which stage the quadrupolar condensate density modulation in the plane perpendicular to the vortex axis induces squeezing in the vicinity of the vortex core. The superflow then undergoes an elliptical instability, the vortex line length stretches and its shape deforms from straight to sinusoidal, oscillating in the plane of the elliptical deformation as illustrated in Fig.\ref{fig1}(b). The elliptical instability explains how the two-dimensional perturbation yields three-dimensional flow instability, and it is driven by a parametric resonance between the $\ell =-2$ surfon and a pair of $\ell =-1$ kelvons with opposite axial momenta. The elliptical instability mechanism has been reviewed by Kerswell in the context of classical hydrodynamics \cite{Kerswell2002a}. As the instability grows the shape of the vortex changes from plane-sinusoidal to helical as shown in Fig.\ref{fig1}(c). Both the vortex core radius and the amplitude of these Kelvin waves are larger near the ends of the condensate, explained by the lower local particle density in those regions. These Kelvin waves decay rapidly via phonon emission into kelvons of longer wavelength. The phonons are visible as ripples on the condensate surface in Fig.\ref{fig1}.  After the primary kelvons have decayed the system is left in a bent vortex state, Fig.\ref{fig1}(d), and the final configuration is determined by the amount of angular momentum left in the system after the external perturbation is switched off. If, instead, the co-rotating quadrupole surfon is resonantly populated using $\Omega=0.8\omega_\perp$ in Eq. (\ref{eq1}), the vortex remains straight for all times and the $\ell =+2$ surfon population is undamped as is verified by our simulations. In this case the conserved angular momentum of the system after the $35$ ms excitation is $L_z/N=1.51\hbar$ and the condensate density oscillates with the characteristic quadrupole frequency $f_{\ell =+2}=162$ Hz.

\begin{figure}
\includegraphics[width=\columnwidth]{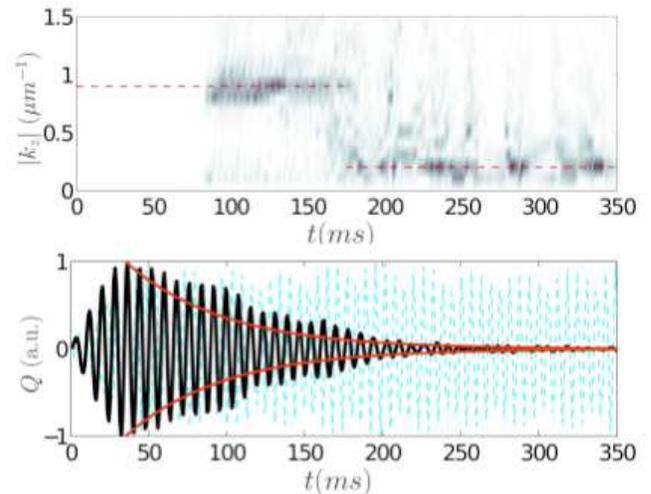}
\caption{(Color online) Kelvin wave excitation and decay (a) and damping of quadrupole modes (b) as functions of time. Frame (a) is obtained by Fourier transforming the azimuthal angle of the vortex position vector as a function of $z$. The quadrupole moments in (b) are computed for $\ell=-2$ (solid line) and $\ell=+2$ (dashed line) perturbations, respectively. The envelope functions are exponentially vanishing with the decay constant $\Gamma=16 s^{-1}$.}
\label{fig2}
\end{figure}

To further quantify the kelvon excitation and decay process, we have located the planar position $(x_v,y_v)$ of the vortex phase singularity as functions of $z$ and $t$. Fourier transform $\mathcal{F}$ of the complex signal $\Theta(z)=\arctan(y_v/x_v)+i\arctan(x_v/y_v)$ reveals clear peaks at wave vectors corresponding to the kelvon excitations present in the system. Figure \ref{fig2}(a) shows $|\mathcal{F}[\Theta(z)]|^2$ as a function of time for the $\ell =-2$ initial perturbation. There emerges a strong signal at $|{\bf k}_z|= 0.9 \mu$m$^{-1}$ which shifts rapidly around 150 ms to $|{\bf k}_z'|= 0.2 \mu$m$^{-1}$. Figure \ref{fig2}(b) shows the normalized quadrupole moment $Q(t)=\int xy n({\bf r}) \; d{\bf r}$, where $n({\bf r})$ is the condensate density, of the system as function of time for the $\ell =-2$ (solid line) and $\ell =+2$ (dashed line) perturbations. The signal for counter-rotated case is strongly damped due to the resonant surfon-kelvon coupling while the co-rotated surfons remain essentially undamped. The exponentially decaying envelope $\propto e^{-\Gamma t'}$, with $\Gamma=16 s^{-1}$ in Fig.\ref{fig2}(b) illustrates the damping of the quadrupole oscillation with the carrier frequency $f_{\ell =-2}=124$ Hz. Landau damping due to the 30\%  thermal fraction in the experiment is likely to explain the faster observed damping rate of $57\pm10 s^{-1}$ \cite{Bretin2003a}.

\begin{figure}
\includegraphics[width=0.9\columnwidth]{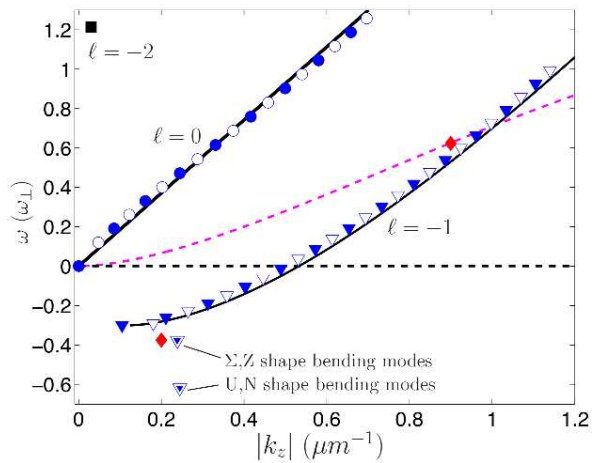}
\caption{(Color online) Kelvin wave dispersion relation showing Bogoliubov excitation frequencies of the $\ell =0$ phonon  (circles)  and $\ell =-1$ kelvon (triangles) modes as functions of the axial momentum. The square at $\omega=1.21 \omega_\perp$ denotes the $\ell =-2$ surfon and the two diamonds near the kelvon dispersion are the initial and final kelvon modes whose wave vectors and energies are obtained from the dynamical simulations. The solid line and curve respectively correspond to the analytical phonon and kelvon dispersion relations in the text. The dashed curve demonstrates the failure of the original kelvon dispersion relation.}
\label{fig3}
\end{figure}

\begin{figure}
\includegraphics[width=0.9\columnwidth]{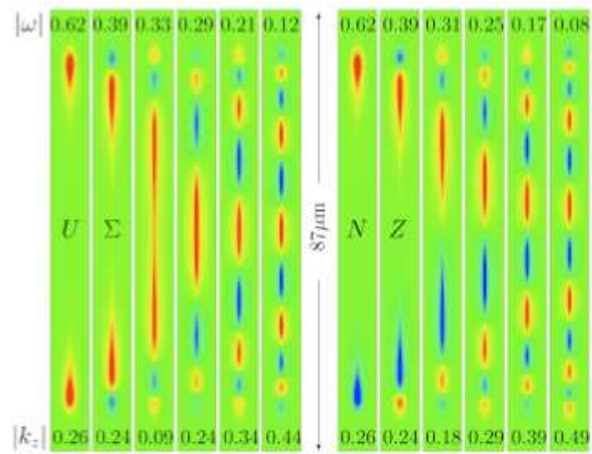}
\caption{(Color online) Bogoliubov kelvon modes for lowest even (left) and odd (right) $z$-parity modes corresponding to those in Fig.\ref{fig3}. The numbers on top and bottom of each frame indicate the mode frequency and momentum, respectively. The color indicates amplitude and sign of the displacement from the equilibrium position}
\label{fig4}
\end{figure}

To complete the picture, we have computed the long-wavelength Kelvin wave dispersion relation by explicitly solving the Bogoliubov-de Gennes equations for the initial state corresponding to Fig.\ref{fig1}(a) \cite{Mizushima2003a,Mizushima2004b,Feder2001a}. Figure \ref{fig3} shows the excitation frequencies for odd (open markers) and even (solid markers) $z$-parity modes as functions of axial momentum for $\ell =-1$ kelvons (triangles) and $\ell =0$ phonons (circles). The straight line is plotted for $\omega_p=c_z|{\bf k}_z|$, where the chemical potential $\mu$ in the speed of sound $c=\sqrt{b\mu/m}$ is renormalized by the factor $b=1/3$ due to the inhomogeneous density. The square  at $\omega=1.21 \omega_\perp$ denotes the counter-rotating $\ell =-2$ quadrupole excitation. This surfon state is initially populated by the resonant ($\Omega=-0.6\omega_\perp$) perturbation. The two lowest energy kelvon modes are the degenerate ``U'' (even parity) and ``N'' (odd parity) shape bending modes \cite{Rosenbusch2002a} while the second lowest modes correspond to the even parity ``$\Sigma$", seen in Fig.\ref{fig1}(d), and odd parity``$Z$" shapes (see also Fig.\ref{fig4}). Dynamically these in-plane bending modes preserve their shape and they precess about the $z$-axis of the trap. The presence of the bending modes in the spectrum manifests the inhomogeneous density due to the trapping. Notice also that the axial momentum is not a good quantum number due to the axial trap potential. Nevertheless, we may assign a well-defined momentum $\hbar{\bf k}_z$ for the Bogoliubov modes\cite{Mizushima2004b}. The two diamonds near the kelvon dispersion are the initial and final kelvon modes, shown respectively in Fig.\ref{fig1}(b) and Fig.\ref{fig1}(d), whose wave vectors and energies (rotation frequencies) are extracted from the time-dependent simulations. The sign of the kelvon frequency determines the sense of rotation of the vortex in the laboratory frame. In Fig.\ref{fig4} we have plotted the Bogoliubov $u_{k}(x,z)$ modes for a few of the lowest collective even (left) and odd (right) $z$-parity kelvon excitations \cite{Mizushima2003a, Mizushima2004b}. The numbers on top and bottom of each frame indicate the excitation frequency and momentum, respectively, of the mode and correspond to those in Fig.\ref{fig3}. All kelvons are highly localized within the vortex core.

The Kelvin wave dispersion relation involving Bessel functions is derived for an infinitely long classical vortex and its long wavelength limit is often applied to superfluid vortices by simply replacing the classical circulation by its quantum mechanical counterpart \cite{Donnelly1991a}. Guided by our numerical experiments for $N$ in the range $(1-50)\times 10^4$, we propose a modification to such semi-classical dispersion relation extending its validity to the finite and inhomogeneous systems:  
\begin{equation}
\omega(k_0+{\bf k}_z ) = \omega_0 +\frac{\hbar {\bf k}^2_z}{2m}  \log \left( \frac{1}{|r_c{\bf k}_z|} \right) , \hspace*{2mm} |r_c{\bf k}_z|\ll 1 
\label{dr}
\end{equation}
where $\omega_0$ is the frequency of the kelvon with the smallest axial momentum $\hbar k_0$ and the system specific vortex core parameter $r_c=0.13\mu$m is only weakly dependent on the density. As shown in Fig. \ref{fig3}, Eq. (\ref{dr}) (solid curve) is in an excellent agreement with the Bogoliubov spectrum (except for the pure bending modes). We have also plotted Eq. (\ref{dr}) using  $\omega_0=k_0=0$ and $r_c=\xi=0.3\mu$m (dashed curve) showing the failure of the original kelvon dispersion relation. We emphasize that the usual density dependent healing length $\xi$ provides poor estimate for the constant $r_c$ for all values of $N$ studied here.

Here we propose a simple method to excite kelvons of specific wave number on demand. We may still make use of the parametric surfon-kelvon resonance by simply shifting the whole kelvon dispersion curve with respect to the quadrupole frequency. This is achieved by adding a Gaussian pinning potential
\begin{equation}
V_{\rm pin}(x,y,z)= \frac{V_0\sigma_0^2}{\sigma(z)^2} \exp \left(\frac{-2[x^2 + y^2] }{\sigma(z)^2}\right)
\label{exppot}
\end{equation}
where $\sigma(z)=\sigma_0\sqrt{1+(z/z_{\rm R})^2}$ and we have chosen $\sigma_0=2\mu$m and a Rayleigh range of $z_{\rm R}=20\mu$m. The vortex pinning effectively modifies the local potential at the vortex core \cite{Isoshima1999a,Simula2002b}. The strength of the applied potential $V_0$ may be used to control the wave vector of the kelvon resonant with the quadrupole surfon. Since the pinning potential is localized in the centre of the trap it does not radically affect the resonant quadrupole surfon frequency itself. To demonstrate the feasibility of this method we have performed simulations using different values for $V_0$ in Eq. (\ref{exppot}) the only difference to the previously described excitation method being the addition of the pinning potential at $t=0$. Setting $V_0 = $\{8.0, 6.0, 4.0, 2.0, -2.0, -4.0\}$\hbar\omega_\perp$ results in an initial kelvon excitation with $|{\bf k}_z|= $\{0.2, 0.4, 0.7, 0.7, 0.9, 1.0\}$\mu$m$^{-1}$, respectively. Experimentally, this method could be realized using tightly focused optical potentials allowing full control over the kelvon excitation process. 

In conclusion, we have theoretically studied the microscopic excitation and decay mechanism of superfluid Kelvin waves in trapped Bose-Einstein condensates.  Our results provide direct verification for the experimental observations of Bretin \emph{et al.} \cite{Bretin2003a}. We interpreted the coupling between the external perturbation and the excitation of Kelvin waves in terms of an elliptical instability mechanism driven by a parametric resonance between the counter-rotating quadrupole surfon and a pair of kelvons of opposite momenta. Subsequently, the Kelvin waves were found to decay to longer wavelength excitations via phonon emission. We presented a modification to the Kelvin wave dispersion relation for inhomogeneous superfluids. Finally, we proposed an experimental method to excite Kelvin waves of any wave number.  We have shown that the pure kelvon decay provides a powerful damping mechanism in superfluid systems even at zero temperature. This may be viewed as an elementary decay channel for superfluid turbulence, the higher order nonlinear processes involving vortex (self) reconnections. Our methods also enable us to go beyond the in-plane Tkachenko modes \cite{Coddington2003a, Baym2003a, Simula2004a, Mizushima2004c, Baksmaty2004a} and to study the full dispersion relation of Kelvin-Tkachenko collective modes in three-dimensional vortex lattices.

\begin{acknowledgments}
We would like to thank N. Hayashi for discussions. This work was supported by the Japan Society for the Promotion of Science (JSPS).
\end{acknowledgments}

\end{document}